\documentclass[12pt]{article}
\usepackage{epsf}
\newcommand{\SMALLCAP} [1]     {\caption[]{\begin{small} #1 \end{small}}}
	
\begin{document}
\bibliographystyle{prsty}
\title{Effects of regulation on a self--organized market}
\author{Gianaurelio Cuniberti$^1$\phantom{\thanks{e-mail: {\tt cunibert@mpipks-dresden.mpg.de}}}, Angelo Valleriani$^2$,  and {Jos\'e Luis Vega}$^3$
\\ $^1${\small {\it Max Planck Institute for the Physics of Complex Systems, D-01187 Dresden, Germany} }
\\ $^2${\small {\it Max Planck Institute of Colloids and Interfaces, Theory Division, D-14424 Potsdam, Germany} }
\\ $^3${\small {\it Banco Bilbao Vizcaya Argentaria, E-28033 Madrid, Spain}} 
}
\date{December 19, 2000}

\maketitle

\begin{abstract}
\noindent
Adapting a simple biological model, we study the effects of
control on the market. Companies are depicted as sites 
on a lattice and labelled by a
fitness parameter (some `company--size'  indicator). The 
chance of survival of
a company on the market at any given time is related 
 to its fitness, its position
on the lattice and on some particular external 
influence, which may be considered to 
represent regulation from governments 
or central
banks. The latter is rendered as a penalty for companies which show a very
fast betterment in fitness space. As a result, we find that the
introduction of regulation on the market contributes to 
lower the average fitness of companies. 
\end{abstract}

\vspace{1.cm}


Complex extended systems showing critical behavior, a lack of
scale in their features, appear to be widespread
in nature, being as diverse as earthquakes \cite{CJ89-GS96},  river
 networks \cite{MCFCB96}, and
biological evolution \cite{BS93}. The
 common characteristic of these systems is that they do
not need any fine tuning of a parameter to be in a critical state.
In an attempt to explain this behavior, Bak, Tang and Wiesenfeld
introduced the concept of self--organized criticality (SOC) \cite{BTW87}. 
In the critical state,   there are
long range interactions, by which 
each part of the system feels the influence of all the others. More precisely,
this means that many of the relevant observables in the system follow
a power--law or Pareto--L{\'e}vy distribution with a non-trivial exponent.

Human society is, by far, one of  the more complex 
extended systems. Its self--organisation is 
evident:
Every single characteristic (historical,
cultural, etc.) of a member of a society influences all 
the other members around him, and through them all the others, 
 in a non--trivial way.
However, despite this complexity, it is universally 
accepted that  cultural details play an interesting but 
somewhat ``marginal'' role in the human history. 
Economic development has always been considered 
the driving (or relevant) 
force in determining the relationships inside a society (for  new perspectives
on this old topic, the reader is kindly referred  to \cite{Lucas87}).

 Moreover, 
it has been confirmed, first in the context of paleontology,  that
evolution and extinction follow a pattern 
 of {\em punctuated equilibrium} \cite{Raup86-EG72-GE77-EG88}.
In a nutshell,  punctuated equilibrium means that long periods of
stasis, during which no significant changes take place, 
are interrupted by sudden bursts of activity,  
that may last for a very short time, after which the qualitative
structure of the system might be completely changed. 
 Models for biological
evolution have shown how punctuated equilibrium--like behavior 
arises naturally  if the
system is in a  self--organized critical state \cite{BS93}. 

Something similar could be said about human history. Indeed,
 the fact that it follows a ``punctuated'' pattern is a widely
acknowledged observation:
Wars, famines, revolutions and counter--revolutions 
are the most
evident (and extreme) illustrations of 
these bursts of historical activity. 
There is enough data in the last ten years
 (not to speak of the last ten centuries), to illustrate this fact. 
It is then natural, if nothing else by the force of mere analogy, 
 to look for evidences 
of critical behavior in social systems. In this paper
we will concentrate on one particular aspect of social systems, namely
economic processes.
Since economy is the one of the most relevant factors
 in determining the structure of a 
society, a great deal of effort has been devoted in recent years
 to the analysis of economic data.
 From  stock--exchange fluctuations \cite{LM99-MS95-Mantegna91}, 
models of production \cite{BCSW93},
to size distribution of companies \cite{Stanley_ea96}, 
it has finally been shown that
 market economy exhibits properties characteristic of 
 a critical system \cite{Mandelbrot97-MS97-MS99}. 
Although we are still far from being able to predict 
 large--scale phenomena, such as  recessions, crashes, etc.,
these results {\it lend credence to the theory}
 that economy is indeed
a self--organized critical system. Bearing these 
considerations in mind, 
we present here a simple model that shows  SOC on
a macroeconomic level together with some of its implications.

In particular, we consider a modified version of
 the model proposed by Bak and 
Sneppen (BS)  to describe the co--evolution
of natural species \cite{BS93}.
 In our model, an economical system (a market) 
will be described by a one--dimensional lattice, 
every site of
which represents  a company or a production sector. 
Companies with stronger mutual 
interactions (because they offer the same product 
or one furnishes goods
to the other)
are arranged on  neighboring sites
(this is similar to the food--chain in the biological case).
 Each company is characterized by a number (its fitness)
describing, for example,  the size of  the company, its  average
earnings or any suitable combination of both. In more general terms, 
 the fitness represents the chance 
a company has to survive in the market. 
As we shall see in the context of our model, 
this represents only the microscopic point of view. Indeed, from a   
macroscopic point of view, any company with a fitness 
beyond a certain threshold survives with the same probability. 
Thus, the company with the lowest fitness is the one that feels 
the strongest pressure to change. 
One can then say that, formulated in these terms, 
the fitness carries global information about, and over to, all
the market chain. As an example, the company with the lowest fitness may 
 be one  working  in a rapidly expanding sector or in a sector 
that, for some reason,  cannot develop any further.
 However, on a macroeconomic level
we do not need to explain which sector or which company might be 
involved or why.  Indeed,  at any 
given time, there are production or distribution sectors 
which are developing or shrinking and companies that 
are in a key position in those sectors. Moreover, on a long time basis,
 it is rather 
hard to predict which sector will 
be the most active  and what will be the consequences of that on
a microeconomic level. Correspondingly, in the simplified situation 
 we are considering,   we neglect the distinction among different specialisations: This is
justified by the observation that, on a macroeconomic level, all different
branches are necessary and in mutual interaction \cite{Bryant88}.

The fate of the worst fit company is either to evolve or go bankrupt 
with its place taken by some new company in the same economical
niche. The nearest neighbour companies will
find a different environment, and their fitnesses will be changed too.  As
a result the system exhibits 
sequences of causally connected evolutionary events called {\em avalanches} \cite{BS93}. An avalanche has a duration $s$ distributed as follows
\begin{equation}
\label{ava}
Q(s)\sim s^{-\tau}\, ,
\end{equation}
where $s$ is the duration of the avalanche and 
$\tau\sim 1.07$ \cite{MPB94-PMB96} 
is the avalanche critical exponent.  
This kind of behavior,
which is the essence of self--organized criticality,
is also the one responsible for the above--mentioned 
punctuated equilibrium.

In the original (biological) BS model, the mutation of the least fit species 
is realized by giving to it
a new fitness taken at random \cite{BS93} from a uniform distribution.
More precisely, each lattice site $j$ is assigned a
random number $f_j$ between $0$
and $1$.  At each time step the smallest
$f_{\rm min}=\min_j\{f_j\}$ is found.  Then, $f_{\rm min}$ and the two nearest
neighbors are updated according to the rule
\begin{equation}
f^{\rm new}=F(f^{\rm old}) \label{eq:noise}
\end{equation}
that assigns a new fitness $f^{\rm new}$ to the chosen lattice site.
 The function $F$ is just a random function
with a uniform distribution between 0 and 1. 
After a transient time, 
the system reaches a stationary critical state in which the distribution of
fitnesses is zero below a certain 
threshold $f_{\rm c}\sim 0.66702(8)$ \cite{MPB94-PMB96} and uniform above it. 
 It has also been shown that the updating rule (\ref{eq:noise}) need not be 
 a random function. Indeed, deterministic updating has also been
considered \cite{DVV97,DVV98} and,  provided the updating rule 
has no long--time  correlations, SOC is preserved together with  
the universality class. 

There is, however, a very important difference between an ecosystem and an economic
market. In a competitive market, a company does not change at random. Quite on
 the contrary,   it will always try to increase its fitness, {\em i.e.} to 
move ``upwards''  towards higher efficiency 
 in order to improve  its chances to stay in business. 
 Thus,  when modelling  the change the company makes to survive
in the market, one has to include a deterministic component, while 
still keeping the ``random character'' of the market \cite{Cootner64}. 
The introduction of chaotic (deterministic) instead of random
updatings take into account both the rather unpredictable variation of the 
market and the voluntary moves of the head of the company.  
Indeed, according to \cite{DVV98}, chaotic (deterministic) updatings 
do not change the statistical properties of the system. 
 

 One very important difference between markets and ecosystems 
still requires consideration: 
 For a given company, too  rapid a growth might 
be a source of problems in the very near future. Indeed, a
very rapidly growing company may face a series of new duties (from new tax
rates to new delivery and/or new advertising systems or new
very expensive technology) that its structures are not able to support. In
other words, 
the market rules may be different 
according to the size of the company.
Adaptation to new rules/tasks might require time,  and when time is
not enough this may lead to bankruptcy.
From this point of view, we are looking for a version of the BS
model which still shows self--organized criticality but which nevertheless
includes some instability for those companies that have experienced too big an 
increase in their fitness. 

For instance, let us suppose that the parameter $f$ represents the 
size (suitably normalized to 1). Let us consider  
the simplified situation in which there are two taxation rates: Companies
whose size is bigger than $\eta$ have to pay a bigger rate than 
those whose size is smaller.
From this point of view, if a company changes its class from  small to
big size, it may face a higher taxation rate that prevents it from 
consolidating its new position. 
We then implement a dynamic rule in the following way: After having assigned
to each site $j$ a random value $f_j$
we proceeds in four basic steps:
\begin{enumerate}
\item{} Find the site of the absolute minimum on the lattice (this site
will be called the active site) and those of the two nearest neighbors.
\item{} Change at the same time their values
by giving them new  numbers according to the given map $F$. 
\item {} If after applying 2, some of the three $f_j$ moves 
from the lower to the upper sector,  we apply  again the map on it. 
\item{} Go to step 1.
\end{enumerate}
Step 3 is tantamount to saying that when a company changes
from the category of ``small business'' to the category of 
``big business'', the rules are different and it has to re--adapt 
to its different environment. Clearly, this step can be accommodated
to  consider any possible 
interpretation of the microscopic parameter $f_j$ at
any given time--step. 
Thus, the dynamic rule we have introduced
can be considered as being a general, if anything a little 
abstract, way of saying that too fast 
a betterment may turn into its contrary.
\begin{figure}[t]
\centerline{\epsfysize=8cm\epsfbox{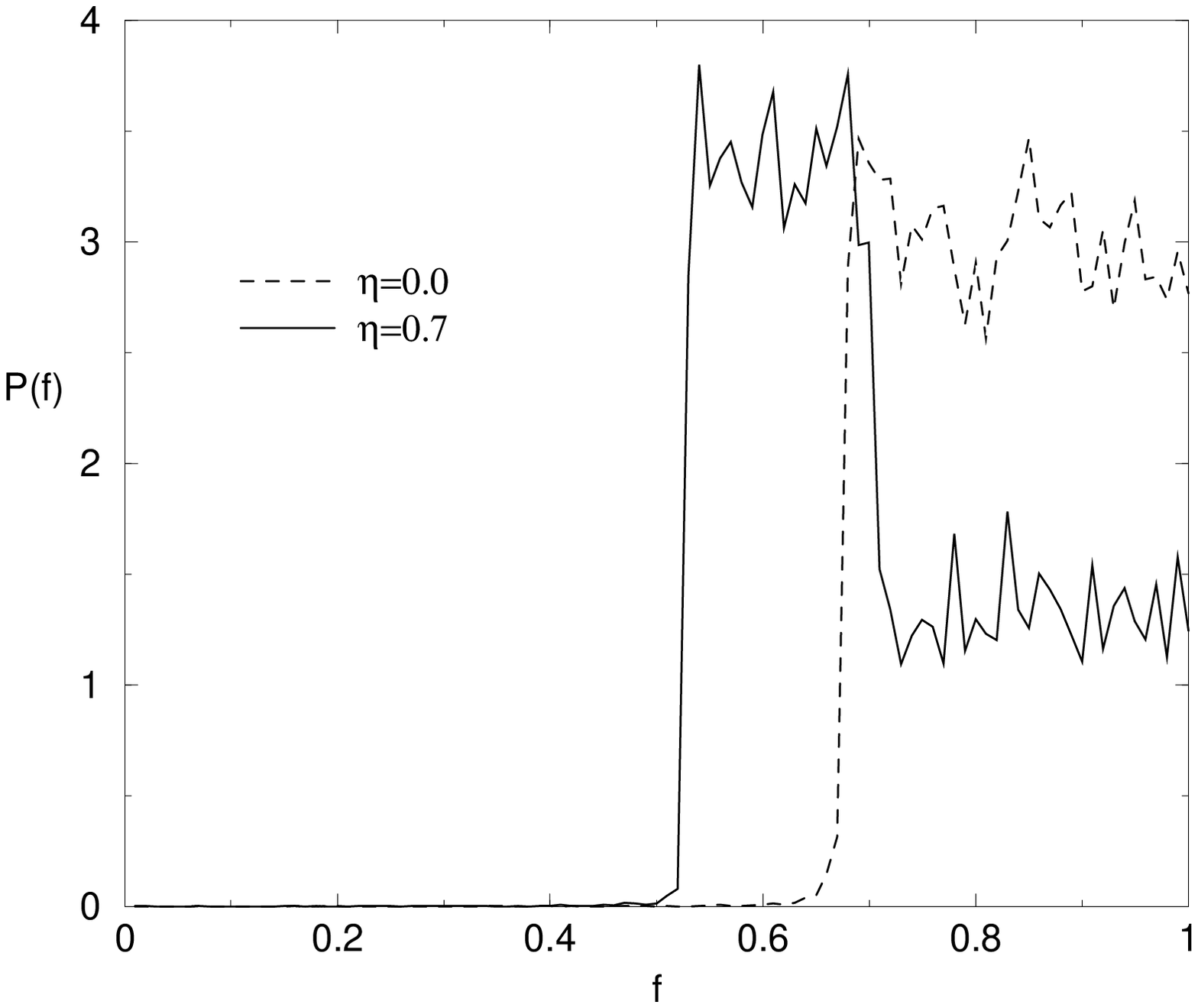}}
\SMALLCAP{Probability density distribution of the fitness parameter $f$ (see text). 
In solid line the distribution shows the effect of
regulations while the dashed line distribution corresponds to the deregulated
market. The non--uniformity comes from the fact that the invariant measure of
the applied chaotic map is also non--uniform.
All simulations were performed on a lattice with 5~000 sites.
\label{fitns_ec}}
\end{figure}

With the rules explained above, the universality class is preserved
and the stationary distribution of 
the $f_j$'s follows also a  pattern similar to that of the random updating
(see Fig.~\ref{fitns_ec}).

The difference between applying or not step 3  
can be observed in Fig.~\ref{fitns_ec}.
The ``control'' applied on the market, namely the 
re--adaptation to different (size dependent) rules,
helps companies of lesser fitness
 to survive. In a sense,
this means that controlling the market helps more
 companies to 
stay in business at the expense of reducing the ``average'' fitness.
 In fact the average fitness goes like 
$\left\langle{f}\right\rangle \simeq (1+f_{\rm c})/2$
and since the threshold decreases, the average fitness
decreases also.

\begin{figure}[t]
\centerline{\epsfysize=8cm\epsfbox{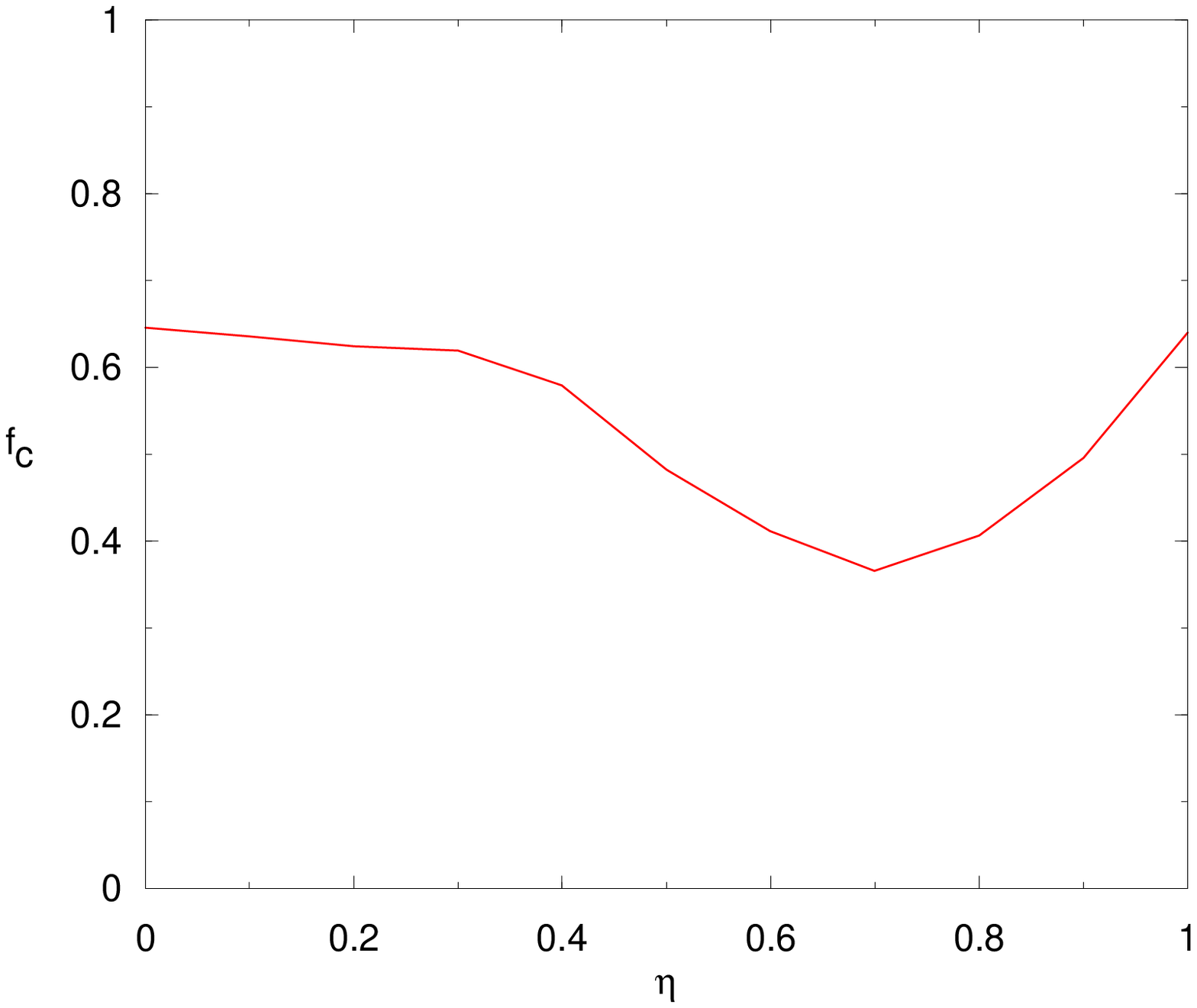}}
\SMALLCAP{The dependence of the critical fitness $f_{\rm c}$ on the value $\eta$
separating taxed and non taxed sectors.\label{fcvseta}}
\end{figure}

Fig.~\ref{fcvseta} shows the resulting dependence of the critical fitness
$f_{\rm c}$ on the value $\eta$,
(according to rules 1--4). 
Notice that in the curve, there is a particular value of $\eta$ where a
minimum occurs. 
This means that slight variations of the parameter $\eta$ do not 
influence the average and critical value of the fitness. 
On the other hand, when $\eta$ is sensibly away from  its minimum, but still 
far from its extreme values, any tiny variation of the taxation levels has amplified
effects on the average fitness.  
In particular it should be noticed that, although the 
boundary cases $\eta=0$ and  $\eta=1$ lead to the same critical 
fitness, in a neighbourhood of them the derivative of the curve is different.  
Tiny variations around $\eta=0$ have smaller effects on the average 
fitness  than those of  tiny variations around $\eta=1$.

The results presented here indicate that, on a global scale, the
behavior of the market is not affected by the 
{\em exact details} of the control rules. On the other hand,  
  time correlations in the chaotic 
map used for the updating  can push the
threshold towards bigger values \cite{DVV98}.
These time--correlations  take into
account how strongly $f^{\rm new}$ is correlated to $f^{\rm old}$ in Eq.\
(\ref{eq:noise}), {\em i.e.} they are related to the decrease in the  average 
randomness in the movements of the variables $f_j$. In other words, an 
on--average stronger control (less ``erratic'' decisions, due for instance to better management) 
on the part of the management of the companies
would have a global consequence, namely an increase of the value of 
the threshold $f_{\rm c}$. In a world where management skills increase with
time (as our world is likely to be),  the threshold $f_{\rm c}$ will also increase
with time and consequently it will be always more difficult to stay in business.

Summarizing, the model introduced in this paper shows which kind of  
changes appear when some control is applied on an SOC system. The exact
nature of the control is, in this simplified model, irrelevant. 
In particular, we show that a control, like the one introduced here, produces always a decrease in the average fitness of the companies operating  in the market. This
could in principle explain why, when deregulated, a market shows initially a
big percentage of companies going out of business. This percentage represents
those ``protected'' by the control that are left unprotected once the control
is removed. 

Caution should be exercised, however, in order to avoid misleading interpretations 
or too simplistic analogies with Darwinian evolution. On the one hand , 
if we look at the fitness parameter, as it is often intended 
in the biological context, one could  infere  that
regulation  subsidizes the less economically adapted companies. On the other hand, as
we mentioned above, the word fitness could refer to 
{\em economical  indicators} --or to functions of them-- related to the size of
a company (income, number of employees, sale size, economic value added\ldots). 
In this sense, a smaller average fitness means a richer variety in the 
economic system.

\end{document}